\documentstyle[12pt]{article}

\begin{document}

\title{Simple Quantum Error Correcting Codes}
\author{A. M. Steane}
\date{\today}
\maketitle
\begin{abstract}
Methods of finding good quantum error correcting codes are discussed, and 
many example codes are presented. The recipe $C_2^{\perp} \subseteq C_1$, 
where $C_1$ and $C_2$ are classical codes, is used to obtain codes for up to 
$16$ information qubits with correction of small numbers of errors. The 
results are tabulated. More efficient codes are obtained by allowing $C_1$ 
to have reduced distance, and introducing sign changes among the code words 
in a systematic manner. This systematic approach leads to single-error 
correcting codes for 3, 4 and 5 information qubits with block lengths of 8, 
10 and 11 qubits respectively. 
  \end{abstract} 

PACS numbers 03.65.Bz,89.70.+c

Two recent papers have shown that efficient quantum error-correcting codes 
exist \cite{Cal96,Ste96}. A quantum error-correcting code is a method of 
storing or transmitting $K$ bits of quantum information using $n > K$ 
qubits\cite{Schu}, in such a way that if an arbitrary subset of the $n$ 
qubits undergoes arbitrary errors, the transmitted quantum information can 
nevertheless be recovered exactly. The only condition is that the values of 
$n$ and $K$, and the structure of the error-correcting code, place an upper 
limit on the number of qubits which can undergo errors before quantum 
information is irretrievably lost. In this context an `error' is any 
physical process which influences the quantum information but whose effect 
cannot be `undone' simply by applying, upon reception, a time-reversed 
version of the `error' process. In practice this will be because the errors 
are unpredictable (eg caused by unknown stray fields) or they entangle the 
information-bearing system with another system which is not accessible to 
detailed manipulation (eg the environment). The latter case, entanglement 
with the environment, includes as a subset relaxation processes such as 
spontaneous emission, and phase decoherence. 

A simple such error correcting code was presented in \cite{Shor}, and a 
general method of encoding a single qubit with correction of
multiple errors was presented in \cite{StePRL}. 
More importantly, further work \cite{Cal96,Ste96} derived whole classes of 
codes, for multiple-correction of many qubits, and showed that {\em 
efficient} codes exist. The word `efficient' in this context refers to the 
fact that the ratio $K/n$, which is called the rate of the code, need not 
become smaller and smaller as $K$ increases, for a given probability of 
error per qubit. However, whereas the pioneering works just mentioned 
established the possibility of efficient quantum error correction, and 
presented the simplest codes possible, they did not address the more 
pragmatic issue of identifying other specific useful codes. This is the 
subject addressed in this paper. 

It is important to identify codes for more than a single qubit, since it is 
known that codes involving more information can be more powerful than simple 
repetition of single-qubit codes. In comparing two coding techniques, one 
may appear more powerful because it can encode a single qubit more 
efficiently. However, if it cannot also be applied to many qubits in an 
efficient way, for example because finding a good code is too difficult, 
then a simpler technique for which multiple-qubit codes can can be found may 
end up being the better choice. 

The earlier simple codes have now been improved upon with the discovery 
\cite{Laf,Ben} of a `perfect' code, that is, one which fills a lower bound 
(elucidated below) on the number of qubits $n$ required to do the job. This 
code is more efficient than those constructed by the recipe of \cite{Cal96} 
and \cite{Ste96}, and introduces an important new class of codes. However, 
it is not yet known how to generalise the construction method in order to 
obtain other efficient codes, so once again the task of identifying specific 
examples of useful codes is an important one. These more efficient codes are 
also discussed here. Simple methods to find good codes are described, and 
three new examples presented.

Note that the errors which we wish to correct are completely random, and
we have no knowledge of their nature other than that they affect different
qubits independently. If we are in possession of further information
about an error process, this can be used to construct codes which are
more resistant to the errors caused by that process. An example is when
the dominant error process is spontaneous emission. In this case
the `error' process is in fact almost completely known, but causes
an unavoidable coupling to the environment. Efficient coding for
this situation has recently been considered \cite{Chuang,Plenio}.
In the present work we make the standard assumption
that the only predictable feature of the errors is their random nature,
so we wish the code to correct as many arbitrary errors as possible
using as few qubits as possible.

Arbitrary errors of qubits can usefully be divided into `amplitude errors', 
that is, changes of the form $\left| 0 \right> \leftrightarrow \left| 1 
\right>$, and `phase errors', that is, changes of the form $\left| 0 \right> 
+ \left| 1 \right> \leftrightarrow \left| 0 \right> - \left| 1 \right>$. 
This division is not meant to imply that these simple state rotations in 
Hilbert space are the only form of error considered, but rather a completely 
general error can be described as a combination of such amplitude and phase 
errors, with associated entanglement with the environment. Less obviously, 
but importantly, a method which can correct both amplitude and phase errors 
is sufficient to correct general errors \cite{Cal96,Ste96,Ekert96}. 

An essential result which relates the problem of error-correction coding of 
a quantum channel to that of a classical channel is embodied in theorems 5 
and 6 of \cite{Ste96} and theorem 1 of \cite{Cal96}: 
  \begin{quote}
{\bf Theorem} (`Quantum Correction Theorem:') {\em If $C_1$ and $C_2$ are 
both linear $[n,k,d]$ codes with $\{0\} \subseteq C_2^{\perp} \subseteq C_1 
\subseteq F_2^n$, then the quantum code $Q_{C_1,C_2}$ is a $t$-error 
correcting code, where $t=\lfloor(d-1)/2 \rfloor$.} 
  \end{quote}
This statement is taken from \cite{Cal96}. Note, however, that we have 
defined $k$ to be the dimension of $C_1$ rather than of its dual as in 
\cite{Cal96}, and replaced $C_2$ by $C_2^{\perp}$, to make the reasoning 
more symetric. The symbol $F_2^n$ refers to the $n$-dimensional vector space 
over a binary field. The phrase `$t$-error correcting code' refers to the 
fact that this form of encoding allows correction after arbitrary errors of 
$t$ qubits. This is stated and proved as a seperate theorem (theorem 6) in 
\cite{Ste96}. In \cite{Ste96}, the equivalent statement of the Quantum 
Correction Theorem is: 
  \begin{quote}
{\em To encode $K$ qubits with minimum distance $d_1$ in one basis, and 
minimum distance $d_2$ in the other, it is sufficient to find a linear code 
of minimum distance $d_1$, whose $K$'th order subcode is the dual of a 
distance $d_2$ code.} 
  \end{quote}

Since $C_2^{\perp}$ has dimension $n-k$, and $C_2^{\perp} \subseteq C_1$, 
clearly $C_2^{\perp}$ is a $K$'th order subcode of $C_1$, with $K = 2k-n$, 
which links the first statement to the second. In the second statement a 
more general form is considered, in which the minimum distances of $C_1$ and 
$C_2$ need not be equal. This has consequences for the type of error which 
can be corrected, as elucidated in \cite{Ste96}. In particular, if only 
phase errors are present, then $d_1$ can be 1, that is, $C_1$ is simply the 
set of all words, so the coding problem reduces to that of finding $C_2 = 
[n,K,d_2]$, which means it is equivalent to the classical coding problem. 
The same simplification applies when only amplitude errors are present, in 
which case $C_2$ is the set of all words and $C_1$ is the error correcting 
code. 

The second statement above refers to two different bases in which the state 
of the information-bearing quantum system may be expressed. For a single 
qubit, basis 1 is $\{ \left| 0  \right>,\; \left| 1  \right> \}$, and basis 
2 is $\{ (\left| 0 \right>+\left| 1 \right>)/\sqrt{2},\; (\left| 0 
\right>-\left| 1 \right>)/\sqrt{2} \}$. For multiple qubits, bases 1 and 2 
are defined to be the respective product bases. The reason why these bases 
are introduced is that correction of amplitude errors is essentially a 
classical error correction operating in basis 1, while that of phase errors 
is essentially a classical error correction operating in basis 2 
\cite{Ste96}. 

The rest of this paper is concerned with finding quantum error-correcting 
codes. We begin with codes obtained according to the recipe of the Quantum 
Correction Theorem. In section \ref{s:0} we consider arbitrary $d_1$ and 
$d_2$, for the case $K=0$. This is not a fruitless exercise since it will be 
shown that a code with $K>0$ can be obtained from one with $K=0$. In section 
\ref{s:K} the case $d_1 = d_2 \equiv d$ is considered, for arbitrary $d$ and 
$K$. This case is important because a completely general error of less than 
$d/2$ qubits can be corrected by the use of such a code (Quantum Correction 
Theorem above, and theorem 6 of \cite{Ste96}). As well as specific codes, 
simple methods for finding codes and for deducing one code from another are 
given. 

Note that whereas the method of sections \ref{s:0} and \ref{s:K} will 
produce a set of useful codes, whose rate does not reduce as $K$ is 
increased at constant error probability per qubit, it will not produce the 
most efficient codes possible. However, the codes may be regarded as a 
starting point from which more efficient codes can be derived by judicious 
use of sign changes with code augmentation (adding of code words) or 
puncturing (deletion of bits). Such methods are discussed in section 
\ref{s:gen}, where optimal or almost optimal quantum codes are presented for 
encoding 3 to 5 qubits with single-error correction. 

Many powerful mathematical (group theoretical) techniques have been applied 
in the pursuit of classical coding theory \cite{MacW}, and since completing 
the present work I have learned of two studies which apply such techniques 
with much success to quantum coding \cite{Cald96b,Gott96}. In this paper, 
the approach is to use simple concepts such as Hamming distance, parity 
check and generator matrices, and examine methods to convert classical codes 
into good quantum codes. The simplicity of these concepts has the advantage 
of being suggestive of useful coding techniques, since they make the 
structure of the codes simple to appreciate, but they do not always lead to 
analytical proofs of the properties of the quantum codes, for which one must 
resort to computer testing. The two approaches of trying simple ideas and 
applying powerful analytical methods are both useful in the quest to find 
good codes. The matrix methods used here have the further advantage that the 
quantum networks for encoding and correction can be derived quite 
straightforwardly from the parity check and generator matrices, by 
generalising the method described in \cite{Ste96}. 

To distinguish the various types of code, it is helpful to have a concise 
notation. Classical linear error correcting codes are identified by the 
notation $[n, k, d]$, meaning a code by which $n$ classical bits can store 
$k$ bits of classical information with minimum distance $d$, hence allowing 
correction of up to $\lfloor (d-1)/2\rfloor$ errors. The notation 
$\{n,K,d_1,d_2\}$ is here introduced to identify a ``quantum code'', meaning 
a code by which $n$ quantum bits can store $K$ bits of quantum information 
and allow correction of up to $\lfloor (d_1-1)/2\rfloor$ amplitude errors, 
and simultaneously up to $\lfloor (d_2-1)/2\rfloor$ phase errors. For codes 
with $d_1 = d_2 \equiv d$ the notation will be abreviated to $\{n,K,d\} 
\equiv \{n,K,d,d\}$. Such codes allow recovery after arbitrary error of up 
to $\lfloor (d-1)/2\rfloor$ of the quantum bits. It may be argued that the 
Hamming distance $d$ is no longer a useful term in the quantum context, 
since it is not clear whether $2t+1$ always corresponds to a quantity with 
the correct properties to be called a `distance' between code vectors of a 
quantum $t$-error correcting code. However, I retain the use of $d$, both 
because it implies the distinction between error detection and correction, 
and because the concept of distance remains useful in searching for quantum 
codes, as will be shown in section \ref{s:gen}. 

The recipe $C_1 = [n,k,d_1];\; C_2 = [n,k,d_2];\; C_2^{\perp} \subseteq C_1$ 
of the Quantum Correction Theorem leads to a code construction in which each 
code vector (ie encoded version of a given logical symbol) consists of a 
superposition of words with coefficients equal in sign as well as magnitude 
in one of the bases 1 or 2 (though not in the other). One may therefore 
choose the sign of all coefficients in the superposition to be positive, in 
the chosen basis. A code having this special form (ie all those discussed in 
sections \ref{s:0} and \ref{s:K}) will be indicated by appending a 
superscript `$+$' sign to the notation, ie $\{n,K,d_1,d_2\}^+$. In general, 
by allowing more general code vectors, a code having the same correction 
ability but higher rate can be obtained from one with all-positive signs in 
the original basis. In symbols, 
  \begin{equation}
\{n,K,d_1,d_2\}^+ \Rightarrow \left( \; \{n'<n,\, K,d_1,d_2\}^{ }_{ }
\;\;\; \mbox{and/or} \;\;\; \{n,K'>K,\, d_1,d_2\} \; \right),
\label{compare}  \end{equation}
where the implication sign is used to mean that once the left-hand side
code is known, the right-hand side code can be obtained easily. 

\section{Zero information qubits}  \label{s:0}

This section will consider $\{n,0,d_1,d_2\}^+$ codes. If $d_1$ is the 
minimum distance of a classical linear code $C$, then by the Quantum 
Correction Theorem (second statement above), $d_2$ is just the minimum 
distance of the dual code $C^{\perp}$, when $K=0$. In symbols, 
  \begin{equation}
[n,k,d], d^{\perp} \equiv \{n,0,d,d^{\perp}\}^+.
  \end{equation}

Helgert and Stinaff \cite{HS} have prepared a table of the minimum distance 
$d$ of linear codes $[n,k,d]$ for given $n$ and $k$. Specifically, the 
interesting quantity is the highest $d = d_{\rm max}(n,k)$ permitted for the 
given values of $n$ and $k$. If $d_{\rm max}(n,k)$ is not known then Helgert 
and Stinaff give upper and lower bounds on it. For brevity, Helgert and 
Stinaff's table will be referred to as HS. It is possible to convert such a 
table into one providing a lower bound on the smallest number of bits $n = 
n_{\rm min} (d,d^{\perp})$ necessary in order that a code can have distance 
$d$ and its dual have distance $d^{\perp}$. For a given $d$, one commences 
with $n=d$, which gives a code $[d,1,d],\,d^{\perp}=2$. To allow larger 
values of $d^{\perp}$, $n$ must be increased, and $k$ set to the largest 
value allowing an $[n,k,d]$ code, as indicated by HS. The values of $n$ and 
$k$ are increased together in this way until HS indicates that an 
$[n,n-k,d^{\perp}]$ code is possible. Clearly, there is no code with $n$ 
smaller than the value thus obtained, for which both $[n,k,d]$ and 
$[n,n-k,d^{\perp}]$ codes are possible. This does not prove, however, that 
an $[n,k,d]$ code exists whose dual has distance $d^{\perp}$. A necessary 
but not sufficient existence condition is established, or in other words, a 
lower bound on the value of $n_{\rm min}(d, d^{\perp})$. This lower bound is 
given in table \ref{tab1}. 

To find out whether the lower bound in table \ref{tab1} is sharp, I have 
attempted to identify codes which satisfy the bound. Success at identifying 
such a code is indicated by an underlined $n$ value in table \ref{tab1}, and 
the code identified is described in table \ref{tab2}. An asterisk in table 
\ref{tab1} indicates that a code with $n$ close to the lower bound exists 
and is identified in the caption. 

\begin{table}
\[
\begin{array}{rr|rlrrlll}
 n      &    & \multicolumn{7}{c}{d} \\
        &    &     3         & \;5 &  7 & 9 & 11 & 13 & 15  \\
\hline
        &  3 &\underline{6}  \\
        &  5 &\underline{11} & 16^* \\
        &  7 &\underline{14}&\underline{20}&\underline{22} \\
d^{\perp}& 9 &\underline{20} & 25 & 30 & 34 \\
        & 11 &\underline{23} & 28^* & 33 & 39 & 42^*  \\
        & 13 &\underline{27} & 33 & 38 & 43 & 46 & 52 \\
        & 15 &\underline{30} & 37 & 42 & 47 & 51 & 56 & 60^*
\end{array}
\]
\caption{
Lower bound on $n$ permitting a dual pair of codes of distances $d$ and 
$d^{\perp}$. The underlined figures (the column $d=3$ and the row 
$d^{\perp}=7$) indicate that the bound is sharp and the code is given below. 
(*) A reduced Golay code can be used to obtain $d=d^{\perp}=5$ with $n=18$. 
The $[31,16,5]$ BCH code has $d^{\perp}=12$, so $d,d^{\perp}=5,11$ is 
possible with $n=30$. There are Quadratic Residue self-dual codes with 
parameters $[48,24,12]$ and $[80,40,16]$ so $d=d^{\perp}=11,\,15$ is 
possible with $n=46,\,78$ respectively \protect\cite{MacW}. } \label{tab1} 
\end{table} 

\begin{table}
\[
\begin{array}{rccll}
n &  k  & d & d^{\perp} & \mbox{code } C \\
\hline
2 &  1  & 2   & 2 & \mbox{Repetition} \\
6 &  3  & 3   & 3 & \mbox{Hamming} \\
11&  4  & 3   & 5 & [8,4,4] \mbox{ extended Hamming } + [3,3,1] \\
14& 10  & 3   & 7 & \mbox{Hamming} \\
20& 15  & 3   & 9 & [16,11,4] \mbox{ extended Hamming } + [4,4,1] \\
23& 18  & 3   &11 & [16,11,4] + [7,4,3] \mbox{ Hamming} \\
27& 22  & 3   &13 & [16,11,4] + [8,4,4] + [3,3,1] \\
30& 25  & 3   &15 & \mbox{Hamming} \\
24 & 12 & 8   & 8 & \mbox{Golay} \\
n &  1  & n   & 2 & \mbox{Repetition } \leftrightarrow \mbox{ even weight} \\
2^r - 1 & n-r & 3 & 2^{r-1} & \mbox{Hamming } \leftrightarrow \mbox{ Simplex} \\
2^r - 1 &n-2r\;&5 & 2^{r-1} - 2^{\lfloor r/2 \rfloor} \; &
\mbox{BCH } \leftrightarrow \mbox{ BCH}^{\perp}
\end{array}
\]
\caption{
Properties of codes making up table 1. The size of the code is $k$, that of 
the dual is $n-k$. Where the code is identified as a sum of two or more, the 
first code in the sum is extended by the others in a manner explained in the 
text. Other codes may be possible, having the same $\{n,k,d,d^{\perp}\}^+$ 
but a different structure. However, there are no linear codes of smaller $n$ 
for the same $d, d^{\perp}$, with the exception of the final entry: 2-error 
correcting BCH codes are not necessarily optimal. They are included here 
because they are close to optimal and easily constructed. } \label{tab2} 
\end{table} 

In table \ref{tab2}, the identification $[n,k,d] + [n',k',d']$ refers to a 
code built by combining two others as follows. To the check matrix of the 
first code ($[n,k,d]$) in the sum, additional columns are added as specified 
by the generator matrix of the second code ($[n',k',d']$) in the sum. This 
lengthens the minimum distance of the dual by $d'$ while increasing $n$ by 
$n'$ and reducing $d$. For example, the code identified as ``$[16,11,4]$ 
extended Hamming $+ [4,4,1]$'' is the $[20,15,3]$ code with the following 
check matrix: 
  \begin{equation}
H_1 = \left( \begin{array}{c}
11111111111111110000 \\
10101010101010101000 \\
01100110011001100100 \\
00011110000111100010 \\
00000001111111100001     \end{array} \right)  \label{H20}
  \end{equation}
Its dual has minimum distance $8+1 = 9$.

One can `navigate' around table \ref{tab1} to some extent by use of the 
following two constructions:
\begin{equation}
[n,k,d], d^{\perp} \Rightarrow \left\{
\begin{array}{ll}
\left[n-1, k-1, d\right], & d^{\perp}-1 \\
\left[n-1, k, d-1\right], & d^{\perp}
\end{array}  \right. .      \label{navigate}
\end{equation}
In these two constructions, the code on the right hand side is derived from 
the $C = [n,k,d]$ code on the left hand side by removing a single row from 
the generator ($k \rightarrow k-1$) or parity check ($d \rightarrow d-1$) 
matrix. To see how the minimum distance of the dual code is affected, recall 
that the generator matrix of $C$ is the parity check matrix of $C^{\perp}$, 
therefore deleting a row from the generator matrix of $C$ means deleting a 
row from the check matrix of $C^{\perp}$, and {\em vice versa}. A single row 
deleted from a generator matrix leaves the minimum distance either 
unaffected (the most likely result) or increased. A single row deleted from 
a check matrix leaves the minimum distance either reduced by one (the most 
likely result) or unaffected. 

\section{$K$ information qubits} \label{s:K}

The case $K \neq 0$ will now be addressed. The simplest case to consider is 
that of a classical code which contains its own dual: $ C^{\perp} \subseteq 
C = [n,k,d]$. This is only possible when $2k \ge n$. Such codes have been 
called ``weakly self-dual'' \cite{MacW}. Since $C^{\perp}$ is a subcode of 
$C$, clearly the Quantum Correction Theorem can be satisfied with $d_1 = d_2 
= d$ and $K = 2k - n$, since $C^{\perp}$ is itself the subcode required by 
the theorem. In symbols, 
  \begin{equation}
C^{\perp}\subseteq C=[n,k,d] \;\;\; \Rightarrow \;\;\; C=\{ n, 2k - n, d \}^+.
\label{Kk}
  \end{equation}
In such a case, the error corrector is the same in basis 1 and basis 2. An 
example is the Hamming code discussed in \cite{Ste96,StePRL}. This result 
transforms the search for quantum $\{n,K,d\}^+$ codes to a large extent to a 
search for classical weakly self-dual codes. This was recognised in 
\cite{Cal96}, where a proof was given that weakly self-dual codes exist 
which satisfy the Gilbert-Varshamov bound. However, there exist 
$\{n,K,d\}^+$ codes which cannot be derived from weakly self-dual codes 
(examples are given below), and these can be more efficient (higher $K/n$ 
for given $d/n$) than the best weakly self-dual codes. 

A code contains its dual if and only if all the rows of the parity check 
matrix satisfy all the parity checks (ie $\rm{wt}(H_i \cdot H_j)$ is even, 
for all $i,j = 0 \ldots n-k-1$). This implies that when a single row is 
deleted from the parity check matrix, the resulting code again contains its 
dual. Using the second construction given in (\ref{navigate}), combined with 
equation (\ref{Kk}), one finds 
  \begin{equation}
C^{\perp} \subseteq C = \{ n,K,d\}^+ \;\;\;  \Rightarrow \;\;\;
C'^{\perp} \subseteq C' =  \{n-1,K+1,d-1\}^+.
\label{KK1}  \end{equation}
This allows one to generate codes encoding more quantum information (having 
greater $K$) from ones of smaller $K$, at the expense of reduced $d$. Note 
that $d$ is not {\em required} to fall by 1, but implication (\ref{KK1}) 
states that $d$ does not fall by more than 1 in this construction. 

Next the following question will be addressed: we wish to encode $K$ qubits 
with given $d = d_1 = d_2$. What is the necessary value of $n$? The Quantum 
Correction Theorem  implies that if subcodes of an $[n,k,d]$ code are used, 
then $K = 2k-n$. In the case of single error correction, ie $d=3$, Hamming's 
construction implies 
  \begin{equation}
 k \le n - \left\lceil \log_2(n+1) \right\rceil,
  \end{equation}
therefore, for an $\{n,K,3\}^+$ code,
  \begin{equation}
K \le n - 2\left\lceil \log_2(n+1) \right\rceil.     \label{K11}
  \end{equation}
When $n=2^r-1$ we have a perfect Hamming code, and for this case the code 
contains its dual. Therefore equality holds in (\ref{K11}), and $K = n - 2 
\log_2(n+1)$. The smallest $n$ allowing $d=3$ for values of $K$ in the range 
1 to 16 is indicated in table \ref{tab3}. 

For $K=2$ equation (\ref{K11}) implies $n \ge 10$. In fact $n=10$
is possible using the following code:
  \begin{eqnarray}
H_1 = \left( \begin{array}{c}
1011001000 \\
0101100100 \\
1010110010 \\
0110010001
\end{array} \right), \;\;\;
H_2 &=& \left( \begin{array}{c}
1111001000 \\
0111100100 \\
1010010010 \\
1110110001
 \end{array} \right),   \label{H1H2} \\
D &=& \left(  \begin{array}{c}
0001001100 \\
0000010011
\end{array} \right).
\end{eqnarray}
Here, $H_1$ and $H_2$ give the correctors in bases 1 and 2 respectively, and 
the generator works as follows. Let $C_1$ be the classical code of which 
$H_1$ is the check matrix. The two rows of $D$ are the fourth and sixth rows 
of the generator $G_1$ of $C_1$, which is obtained from the well-known 
relation 
  \begin{equation}
 H_1 = \left(  A \left| I_{n-k} \right. \right)
\Leftrightarrow G_1 = \left( \left. I_{k} \right| A^T \right)
\label{HtoG}  \end{equation}
where $I_j$ is the $j \times j$ identity matrix, and $A$ is the rest of the 
check matrix. Adding these two extra checks to $H_1$, we obtain the check 
matrix for a subcode $C_2^{\perp}$ of $C_1$. The four states (code vectors) 
in the quantum $\{10,2,3\}^+$ code are the subcode $\left| C_2^{\perp} 
\right>$, whose generator is $H_2$, and its three cosets $\left| C_2^{\perp} 
\oplus D_0\right>,\; \left| {C_2^{\perp} \oplus D_1}\right>, \;\left| 
{C_2^{\perp} \oplus D_0 \oplus D_1}\right>$, where $D_0$ and $D_1$ are the 
rows of $D$ (the letter $D$ is chosen here for `displacement'). In symbols, 
one may write this generation procedure as 
  \begin{equation}
G = \left( \begin{array}{c} H_2 \\ \hline D \end{array} \right). \label{GD}
  \end{equation}
This equation may be regarded as a summary of the quantum network which will 
encode the two qubits of information. Note that since the rows of $D$ are 
members of the code $C_1$, they satisfy all the checks of $H_1$, and so the 
cosets they generate are all subsets of $C_1$. Also, since the rows of $D$ 
have odd weight, the coset $\left| {C_2^{\perp} \oplus D_i}\right>$ fails 
the parity check $D_i$, so the cosets are distinct. In general $D$ need not 
have rows of odd weight. The non-overlapping of the cosets is ensured by the 
fact that $C_2$ is not a zero-distance code. 

The matrix formed by $H_1$ plus the extra rows given by $D$ is the generator 
of $C_2$, and the corrector in basis 2, $H_2$, is obtained from this 
generator using relation (\ref{HtoG}). All these relationships may be 
summarised as follows: 
  \begin{eqnarray}
H_1 & \leftrightarrow & G_1 \longrightarrow D, \\
\left( \begin{array}{c} H_1 \\ \hline D \end{array} \right) 
    &\leftrightarrow & H_2,            \label{H1D} \\
H_1 & \leftrightarrow &
\left( \begin{array}{c} H_2 \\ \hline D \end{array} \right) 
    = G.
  \end{eqnarray}
From this one may see that an equivalent code is obtained by using $H_1$ and 
$D$ as the generator in equation (\ref{GD}) instead of $H_2$ and $D$. A 
further equivalent quantum code can be obtained by using the first two rows 
of $G_1$ for $D$, instead of the fourth and sixth rows (cf equations 
(\ref{D12}) and (\ref{G8}) below). 

The above approach can clearly be applied to any classical $[n,k,d]$ code. 
That is, one produces a subcode by using $2k-n$ words from the code as extra 
parity checks, with the aim that the check matrix thus obtained is the 
generator of another (or the same) $[n,k,d]$ code. However, it is not clear 
whether this method can always succeed in producing a useful quantum code. 
For example, whereas the cyclic check matrix $H_1$ of equations (\ref{H1H2}) 
leads quickly to a quantum code, the Hamming check matrix for the same 
parameters $[10,6,3]$ does not lead to a generator in the form given by 
equation (\ref{HtoG}) whose rows can be used to form $D$. It would be 
interesting to try to prove or disprove the hypothesis that the existence of 
a classical $[n,k,d]$ code is sufficient to imply at least the existence of 
a quantum $\{n,2k-n,d\}^+$ code. The author's current impression is that 
this hypothesis is untrue in general. However, it is true for weakly 
self-dual codes, and probably gives a close estimate of the parameter values 
possible for other cases. 

The single-error correcting codes indicated in table \ref{tab3} were all 
obtained by using the above method of using code words as extra parity 
checks, but note that whereas I have thus found single-error correcting 
$\{13,5,3\}^+$ and $\{14,6,3\}^+$ codes, filling the lower limit on $n$ set 
by the Hamming bound, I have not found $\{11,3,3\}^+$ or $\{12,4,3\}^+$ 
codes even though classical $[11,7,3]$ and $[12,8,3]$ codes exist. These 
single error-correcting quantum codes are all obtained from the cyclic code 
given by the irreducible primitive polynomial $x^4 = 1 + x$. The check 
matrix $H_1$ in equations (\ref{H1H2}) is the check matrix of this cyclic 
code for the case $n=10$, and for higher $n$, up to $n=15$, further columns 
are added to the front of the matrix following the standard procedure. Once 
we have $H_1$, the quantum code is fully defined once the relevant 
displacement matrix $D$ is given. For $n=12$ to $14$ the following matrices 
$D$ fulfill the requirements for single error correction: 
  \begin{eqnarray} 
D_{\{12,3,3\}^+} &=& \left( \begin{array}{c} 
001000001010 \\ 
000100000101 \\
000010001011 \end{array} \right), \label{D12} \\ 
D_{\{13,5,3\}^+} &=& \left( 
\begin{array}{c}
1000000001111 \\
0100000001110 \\
0010000000111 \\ 
0001000001010 \\
0000000010011  \end{array} \right), \\
D_{\{14,6,3\}^+} &=& \left( \begin{array}{c} 
10000000001101 \\
01000000001111 \\
00100000001110 \\ 
00010000000111 \\
00001000001010 \\
00000000010011  \end{array} \right). 
  \end{eqnarray}
Note that the cases $n=13$ and $n=14$ are similar to one another, and can be 
obtained by reducing the $n=15$ code. For $n=15$ the code $C_1$ contains its 
own dual. 

Using the above methods, and once again the table of HS, lists of quantum 
$\{n,K,d\}^+$ codes can be compiled. The results are summarised in tables 
\ref{tab3} and \ref{tab4}. The upper bound on $d$ is found from the 
classical bound $d_{\rm max}(n,\,k=(K+n)/2)$ given by HS. As it stands, 
table \ref{tab3} is incomplete in that for most entries I have not found 
codes which realise the upper bound, thus prooving that it is obtainable. 
However, classical self-dual codes supply efficient quantum codes of low 
rate, high distance (low $K/n$, high $d$), and BCH codes supply efficient 
quantum codes of high rate, low distance. Therefore we have identified 
infinite series of codes, as $n$ increases, at the two ends of the range 
$K=0 \ldots \sim (n - \log_2(n))$. 

The $\{17,7,3\}^+$ code in table \ref{tab3} is specified by the following 
check matrix 
  \begin{equation}
H_1 = \left( \begin{array}{c}
01100111100110000 \\
10111100101101000 \\
11010010111100100 \\
11101001110000010 \\
00011111110000001  \end{array} \right)
  \end{equation}
with the $D$ matrix equal to the last 7 rows of $G_1$. This code can be 
obtained by adding a check bit to the $[16,11,4]$ extended Hamming code.

The single-error correcting codes indicated in table \ref{tab3} as certainly 
obtainable (ie which I have succeeded in finding) for $7 < K < 17$ do not 
realise the minimum implied by the results of HS, but require one additional 
qubit, similar to the cases $K=3$ and $4$ already remarked. Using the clue 
mentioned above that a cyclic classical code rather than a Hamming code is a 
good choice, all the codes from $K=8$ to $16$ were obtained from the cyclic 
classical codes of primitive polynomial $x^5 = 1 + x^3$. In this series of 
codes, a classical $[n,k,d]$ code gives rise to a quantum $\{n,2k-n-1,d\}^+$ 
code. For example, the generator for $K=16$ is 
  \begin{equation}
G_{\{27,16,3\}^+} = \left( \begin{array}{c}
011001111100011011101010000 \\
101100111110001101110101000 \\
001111100011011101010000100 \\
100111110001101110101000010 \\
110011111000110111010100001 \\
\hline
001000000000000000000011100 \\
000010000000000000000000111 \\
000000100000000000000011111 \\
000000001000000000000011001 \\
000000000010000000000001100 \\
000000000001000000000000110 \\
000000000000100000000000011 \\
000000000000010000000010101 \\
000000000000001000000011110 \\
000000000000000100000001111 \\
000000000000000010000010011 \\
000000000000000001000011101 \\
000000000000000000100011010 \\
000000000000000000010001101 \\
000000000000000000001010010 \\
000000000000000000000101001
\end{array} \right)
  \end{equation}
The generators for the other codes in this series have a similar form
and will not be listed.

A $\{20,9,3\}^+$ code can also be obtained from the classical code having 
the minimum $n$ for $d=3, d^{\perp}=9$, referred to in tables \ref{tab1} and 
\ref{tab2}. Its check matrix $H_1$ is given in equation (\ref{H20}), and the 
displacement matrix is formed from rows of $G_1$ as follows 
  \begin{equation}
D_{\{20,9,3\}^+} = \left( \begin{array}{c}
01000000000000010100 \\
00000100000000010110 \\
00000001000000010001 \\
00000000010000010101 \\
00000000001000011101 \\
00000000000100010011 \\
00000000000010011011 \\
00000000000001010111 \\
00000000000000111111     \end{array} \right)
  \end{equation}

\begin{table}
\[
\begin{array}{ll|lllllll|llllllll|l}
d&  & \multicolumn{16}{c}{K} \\
 &  & 1& 2& 3& 4& 5& 6& 7& 8& 9&10&11&12&13&14&15&16 \\
\hline
 & 1& \underline{1}^* &&&&&&\\
 & 2& & \underline{1}^*&&&&&\\
 & 3& & & \underline{1}^*&&&& \\
 & 4& \underline{2} & \underline{2}^*& & \underline{1}^* &&&\\
 & 5& & \underline{2}& & & \underline{1}^* &&\\
 & 6^*& & & \underline{2}& \underline{2}^*& & \underline{1}^* &\\
 &7&\underline{3}^*&& & \underline{2}& & & \underline{1}^* \\
\hline
 & 8^*& & & & & \underline{2}& \underline{2}^*& & \underline{1}^* \\
&9&\underline{3}& & & & & \underline{2}& & & \underline{1}^* \\
&10& &\underline{3}& & & & & \underline{2}& \underline{2}^*&
 & \underline{1}^* \\
 &11& 4 & & 3& & & & & \underline{2}& & & \underline{1}^* \\
 &12&  &4&\underline{3}&3& & & & & \underline{2}&
 \underline{2}^*& & \underline{1}^* \\
n&13&  & & 4& &\underline{3}& & & & & \underline{2}&
  & & \underline{1}^* \\
 &14&  & & & 4& &\underline{3}&
  & & & & \underline{2}& \underline{2}^*& & \underline{1}^* \\
 &15&  & & & & 4&&\underline{3}^*&&
 & & & \underline{2}& & & \underline{1}^* \\
\hline
 &16&   & & & &&\underline{4}^*& & & & & & 
& \underline{2}& \underline{2}^*& & \underline{1}^* \\
 &17& 5 & & & & & 4&\underline{3}& & & & & & & \underline{2}& & \\
 &18&   & & & & & 4& & 3& & & & & & & \underline{2}& \underline{2}^*\\
&19&\underline{5}& & & & & & 4&\underline{3}& 3& & & & & & & \underline{2}\\
&20& &\underline{5}&&\underline{4}^*& & & &4&\underline{3}&3& & & & & & \\
&21&\underline{6}&&\underline{5}^*& & & & & & 4&\underline{3}& 3& & & & & \\
&22& &\underline{6}^*&& 5& & & & & & 4&\underline{3}& 3& & & & \\
&23&\underline{7}^*&&6& & 5& & & & & & 4&\underline{3}& 3& & & \\
\hline
 &24^*&  & & & 6& &45& & & & & & 4&\underline{3}& 3& & \\
 &25&  & & & &56& &45& & & & & & 4&\underline{3}& 3& \\
 &26&  & & & 6& &56& &45& & & & & & 4&\underline{3}& 3 \\
 &27& 78& & & & 6& &56& &45& & & & & & 4&\underline{3}  \\
 &28&  &67& & & & 6& &56& &45& & & & & & 4\\
 &29&\underline{6}8& &67& & & & 6& &56& &\underline{4}^*5& & & & & \\
 &30& &\underline{6}^*8& &67& & & & 6& &56&&\underline{4}^*&& & & \\
&31&\underline{7}^*8&&68& &67& & & & 6&&\underline{5}^*&& 4 & & & \\
\hline
\end{array}
\]
\caption{
Upper bound on $d$ for $\{n,K,d\}^+$ codes of small $K$ and $n$. Entries 
which are identical to the one immediately above them (ie with $n$ reduced 
by 1) are left blank, in order to bring out the pattern in the results. A 
pair of figures is given when the table of HS \protect\cite{HS} indicates a 
range of distance values rather than a precise upper limit. The underlined 
values are produced by codes given in table \protect\ref{tab4} or obtained 
from them by the methods discussed in the text. For these codes the listed 
upper bound is thus shown to be obtainable. Some codes are obtained from one 
another by deleting a row of $G_1$ (moving upwards and to the left in the 
table), or by deleting a row of $H_1$ (moving upwards and to the right in 
the table). An asterisk ($^*$) indicates a self-dual or weakly self-dual 
code.} 
  \label{tab3}  \end{table}

\begin{table}
\[
\begin{array}{rcll}
n       &  K  & d & \mbox{code } C^{+K} \\
\hline
4       &   2 & 2 & \\
n       & n-2 & 2 & \mbox{Even weight, for even }n \\
2^{r}-1 & n - 2\log_2(n+1) & 3 & \mbox{Hamming},\;r>2 \\
\mbox{eg. } 7  &  1 & 3 & \\
         15    &  7 & 3 & \\
         31    & 21 & 3 & \\
2^{r}-1 & n - 4\log_2(n+1) & 5 & \mbox{BCH},\;r>4 \\ 
\mbox{eg. } 31 & 11 & 5 & \\
            29 & 11 & 4 & \mbox{reduced BCH,} C^{\perp} \subseteq C \\
2^{r}-1 & n-2 t\log_2(n+1) & 2t+1 & \mbox{$t$-error correcting BCH} \\
\mbox{eg. } 31 &  1 & 7 & \mbox{BCH} \\
16      & 6   & 4 & \mbox{extended Hamming} \\
17      & 7   & 3 & \mbox{see text} \\
19-27   & 8-16 & 3 & \mbox{cyclic, see text} \\
23      & 1   & 7 & \mbox{Golay} \\
\hline
48 & 0 & 12 & \mbox{Quadratic residue (self-dual)} \\
63 & 3 & 11 & \mbox{BCH} \\
63 & 15 & 9 & \mbox{BCH} \\
80 & 0 & 16 & \mbox{Quadratic residue (self-dual)} \\
104 & 0 & 20 & \mbox{Quadratic residue (self-dual)} \\
127 & 15 & 17 & \mbox{BCH}
\end{array}
\]
\caption{
Example $\{n,K,d\}^+$ codes. The final entries give some
assorted values of $n$ and $K$ larger than those covered by
table \protect\ref{tab3}.
}  \label{tab4} \end{table}

\section{More efficient codes} \label{s:gen}

Implication (\ref{compare}) was used in the introduction to encapsulate the 
twin facts that in general $\{n,K,d\}^+$ codes are not the most efficient 
possible, and that they can be used as a starting-point to obtain more 
efficient $\{n,K,d\}$ quantum codes. The simplest example is the perfect 
$\{5,1,3\}$ quantum code described in \cite{Laf,Ben}, which can be obtained 
by deleting any two bits from the $\{7,1,3\}^+$ code described in 
\cite{StePRL,Ste96}, and changing the signs of a subset of the words in each 
of the two code vectors. The relevent sign changes can be found for this 
simplest case by an exhaustive computer search. The computer search is a 
useful tool in the task of finding good codes, which may be likened to a 
search for the best fruit on a many-branched tree. However, a complete 
search of all possible allocations of signs rapidly becomes too time 
consuming, as the parameters $\{n,K,d\}$ are increased. Intelligent search 
techniques must be used, and barren branches of the tree ruled out as 
efficiently as possible, while fruitful branches must be identified before 
the search begins, which is the demanding task of the human researcher. In 
this section a pair of quantum codes will be presented, both of which were 
found by taking advantage of two simple methods to identify fruitful 
branches and thus find suitable sign allocations quickly. Before discussing 
these sign allocations, however, we will consider ways of combining 
classical codes which go beyond the simple recipe $C_2^{\perp} \subseteq 
C_1$. 

The generator matrix of a quantum code, equation (\ref{GD}), creates one 
classical code (forming the first quantum code vector) and $2^K-1$ cosets 
(which form the remaining quantum code vectors). Thus we may picture the 
first quantum code vector as a lattice of points in a $2^n$-dimensional 
Hamming space, and the other code vectors as this lattice displaced around 
Hamming space by distances of order $d$. The codes described in previous 
sections used lattices displaced so that each point in any given lattice was 
at least a distance $d$ from any point in another lattice. In other words, 
the set of all the lattices formed a classical code of distance $d$, and 
this ensured that error correction was possible in basis 1. However, in 
forming a quantum code, it is not necessary to displace the lattices as far 
as this. Bit flips in basis 1, ie amplitude errors, will cause a given 
lattice to move towards some other lattice, ie the code vectors approach, 
but if we now allow the signs in basis 1 to be negative as well as positive, 
then lattices (ie cosets) which overlap, in that they contain the same sets 
of words, may nevertheless correspond to orthogonal quantum states since 
there is an equal number of positive and negative contributions to the inner 
product $\left< i,e_k \right. \left| j,e_l \right>$, where $\left| i,e_k 
\right>, \, \left| j,e_l \right>$ are code vectors affected by errors $e_k, 
e_l$. Thus if we start from a set of code vectors with all-positive signs 
when written in basis 1, then the introduction of sign changes permits the 
distance between cosets in basis 1 to be reduced. 

Clearly, we must not hope for too much from this ability to allow the cosets 
to approach. The minimum assumption is that we may permit the distance 
between cosets in basis 1 to be reduced by one. In other words the recipe 
for a $\{n,K,d\}$ quantum code becomes $C_2^{\perp} \subseteq C_1$ where 
$C_2$ is a distance $d$ classical code as before, but now $C_1$ is a 
distance $d-1$ code. In addition, we wish $C_2^{\perp}$ to have as large a 
minimum distance as possible, in order to allow a lot of `room' to move the 
lattice around in Hamming space before it overlaps itself. These two 
conditions, together with a judicious application of sign changes, will be 
used to find optimal single-error correcting quantum codes. First, however, 
we must consider how to apply sign changes to the words in the code vectors. 

The first method to allocate sign changes to the words in each code vector 
is to restrict the possible sign allocations to those given by rows of the 
Hadamard matrix \cite{MacW}. That is to say, we use the $w \times w$ 
Hadamard matrix to supply $w$ different allocations of $w$ signs. A sign 
allocation is a set of $w$ $+1$'s and $-1$'s, giving the signs of each of 
the $w$ words in the superposition forming the code vector to be tested. The 
Hadamard matrix can be used in this way since since in all the codes 
considered here, each code vector contains a number of words $w$ equal to a 
power of 2. It is not hard to convince oneself that a row of the Hadamard 
matrix is an intelligent choice of sign allocation for any quantum code 
derived from a linear classical code by the methods discussed in previous 
sections (with one or more bits deleted). 

Once we have a code vector, that is, a set of words with a proposed sign 
allocation, it is tested. The test consists of first testing whether errors 
in the code vector lead to states orthogonal to the code vector itself and 
to each other, and then testing whether such erroneous states are also 
orthogonal to all the other code vectors in the code and their erroneous 
versions. The possible errors included in the test are all those which the 
code is supposed to be able to correct. If a code survives such a test, then 
errors of different syndrome lead to orthogonal states, and the 
orthogonality of different code vectors is also preserved. Such an 
`orthogonal coding' implies that error correction is certainly possible. 
This latter fact is a central part of the argument presented in \cite{Cal96} 
and \cite{Ste96}.  An elegant presentation of it is also provided by Ekert 
and Macchiavello \cite{Ekert96}, which enables the latter authors to deduce 
a quantum version of the Hamming bound, based on counting the number of 
possible orthogonal directions in Hilbert space. Their bound is 
  \begin{equation}
2^K \sum_{i=0}^{t} 3^i \left( \begin{array}{cc}
n \\ i \end{array} \right)  \leq 2^n     \label{bound}
  \end{equation}
for the size $K$ of a possible quantum code which can correct $t$ general 
errors using $n$ qubits. This bound is more general than that required for 
$\{n,K,2t+1\}^+$ codes derived in \cite{Ste96}, since it includes the 
possibility of the more general $\{n,K,2t+1\}$ codes which we are 
considering in this section. However, there is an intriguing possibility 
that it is not strictly necessary for all possible error syndromes of all 
possible code vectors to be associated with mutually orthogonal states, 
since some error correction techniques may be able to correct errors of 
different syndrome without needing to distinguish the syndromes explicitly. 
This is the subject of active research \cite{Ben,ShorSmol,Cald96b} and will 
not be addressed in the present work. In other words, we use `orthogonal 
coding' throughout. 

The second of our two methods to identify fruitful trial codes is to 
consider the sign allocation, that is the row of $w$ $+1$'s and $-1$'s, as 
itself a binary vector of length $w$, and then to use linear combinations of 
such vectors in a fashion to be explained shortly. To keep the notation 
concise, we replace $+1$ in the sign allocation vector by $0$, and $-1$ by 
$1$, to get a vector in the usual binary form, but one which is understood 
to represent a sign allocation among $w$ superposed words. Note that this 
vector has length $w$, which is a power of two and usually larger than the 
length $n$ of the words in the superposition forming a code vector. For 
example, the sign allocation for the two code vectors of the five-qubit code 
of \cite{Laf} can be written 
  \begin{equation}  
00010100, 01110010            \label{sign5}
  \end{equation}
where the least significant (ie rightmost) bit in the sign vector gives the 
sign of the first word in the code vector, and we assume the order of the 
words in the code vectors is that obtained when they are generated using the 
generator matrix 
  \begin{equation}
G = \left( \begin{array}{c}
10101 \\
10011 \\
01111 \\
\hline
11111
\end{array}   \right)
  \end{equation}
(cf equation \ref{GD} for the notation). Note that the signs in 
(\ref{sign5}) are {\em not} rows of the Hadamard matrix, showing that the 
Hadamard method will not pick up all good codes. However, both sign vectors 
in (\ref{sign5}) are offset from rows of the Hadamard matrix by the same 
code vector $00010100$, so (\ref{sign5}) is a coset of a sign allocation 
obtained by the Hadamard method (ie a coset of a subset of a first-order 
Reed-Muller code). 

The second of our two methods to allocate signs only applies to codes of 
more than one encoded qubit, ie having more than two code vectors. The 
method is to let the set of $2^K$ $w$-bit sign vectors itself be a classical 
linear code (or a coset of a linear code if necessary), and to allocate each 
sign vector thus generated to the corresponding code vector generated by 
$G$. For example, once we have found sign vectors $s_{00}, s_{01}, s_{10}$ 
for the first three code vectors of an $\{n,2,d\}$ code, we try the sign 
vector $s_{11} = s_{00} \oplus s_{01} \oplus s_{10}$ for the fourth code 
vector. By this process, we only need $K$ sign vectors (plus possibly one 
more to form a coset) to specify all the signs for an $\{n,K,d\}$ code, 
rather than finding $2^K$ independent vectors which is a much more demanding 
task. 

The quantum Hamming bound (\ref{bound}) states that for single error 
correction ($t=1, \;d=3$), at least $n=5,7,8,9,10$ qubits are required to 
encode $K=1,2,3,4,5$ qubits respectively, and $n=10$ qubits are required to 
correct $K=1$ qubit with double error correction ($t=2,\;d=5$). The $n=5$ 
case is a perfect code since it fills the bound, and is that discussed in 
\cite{Laf,Ben}. The next most simple case is $n=7$, for which we search for 
an encoding of 2 qubits with single error correction. I have not found such 
a $\{7,2,3\}$ code, despite a wide but not complete search (this search was 
not restricted to the two methods just discussed). The best codes I have 
found are ones which encode 2 qubits using 7 but for which the third and 
fourth code vectors are not quite compatible with the first and second. That 
is to say, there are 10 cases in which a single-qubit error in one code 
vector leads to the same quantum state as a different single-qubit error in 
another code vector, causing an ambiguity for any error corrector. These 10 
cases are taken out of the $88^2/2 = 3872$ possible comparisons between one 
code vector with its erroneous versions and another code vector with its 
erroneous versions, so the code comes close to single error correction, 
while not realising it completely. 

An encoding of three qubits permitting complete single error correction can 
be obtained with $n=8$, which is optimal in that this is the lower limit 
given by (\ref{bound}). To find the code, we begin with a classical code 
$C_2$ having minimum distance at least 3, to allow correction of errors in 
basis 2 (phase errors) and having a dual $C_2^{\perp}$ of minimum distance 
as large as possible, since this dual code defines the lattice in basis 1 
whose various displaced versions constitute the code vectors in basis 1. The 
$[8,4,3]$ Hamming code is not a good choice since its dual has a minimum 
distance of only 1. Instead we adopt the extended Hamming code or 
Reed-Muller code $C_2 = [8,4,4]$, which is self-dual so $C_2^{\perp}$ has 
minimum distance $4$. Since we want $d=3$, we allow $C_1$ to have distance 
$d-1 = 2$. This suggests the even-weight $[8,7,2]$ code, which has the 
correct number of code vectors to allow $K=3$. Thus we obtain the following 
generator: 
  \begin{equation}
G_{\{8,2,3\}} = \left( \begin{array}{c}
01010101 \\
00110011 \\
00001111 \\
11111111 \\
\hline
11000000 \\
10100000 \\
10001000  \end{array} \right)   \label{G8}
  \end{equation}
The sign vectors are found by computer search using the two short-cuts 
described above, which leads to eight sign vectors generated by 
  \begin{equation}
S_{\{8,3,3\}} = \left( \begin{array}{c}
0011001100110011 \\
0000111100001111 \\ 
0110011001100110 \end{array} \right)
\equiv \left( \begin{array}{c} 3333 \\ 0{\rm F}0{\rm F} \\
6666 \end{array} \right),
  \label{S8} \end{equation}
where the second version is the first written in hexadecimal to bring out the 
structure.

Equations (\ref{G8}) and (\ref{S8}) are quite concise and combine several new 
notations introduced in this paper. To make sure the notation is correctly 
understood, the $\{8,2,3\}$ code defined by these equations is now written out 
in full: 
  \begin{eqnarray*}
\left| {v000} \right>&=& 
+ \left| 00000000 \right>
+ \left| 01010101 \right>
+ \left| 00110011 \right>
+ \left| 01100110 \right> \\ &&
+ \left| 00001111 \right>
+ \left| 01011010 \right>
+ \left| 00111100 \right>
+ \left| 01101001 \right> \\ &&
+ \left| 11111111 \right>
+ \left| 10101010 \right>
+ \left| 11001100 \right>
+ \left| 10011001 \right> \\ &&
+ \left| 11110000 \right>
+ \left| 10100101 \right>
+ \left| 11000011 \right>
+ \left| 10010110 \right>
\\
\left| {v001} \right>&=& 
- \left| 11000000 \right>
- \left| 10010101 \right>
+ \left| 11110011 \right>
+ \left| 10100110 \right> \\ &&
- \left| 11001111 \right>
- \left| 10011010 \right>
+ \left| 11111100 \right>
+ \left| 10101001 \right> \\ &&
- \left| 00111111 \right>
- \left| 01101010 \right>
+ \left| 00001100 \right>
+ \left| 01011001 \right> \\ &&
- \left| 00110000 \right>
- \left| 01100101 \right>
+ \left| 00000011 \right>
+ \left| 01010110 \right>
\\
\left| {v010} \right>&=& 
- \left| 10100000 \right>
- \left| 11110101 \right>
- \left| 10010011 \right>
- \left| 11000110 \right> \\ &&
+ \left| 10101111 \right>
+ \left| 11111010 \right>
+ \left| 10011100 \right>
+ \left| 11001001 \right> \\ &&
- \left| 01011111 \right>
- \left| 00001010 \right>
- \left| 01101100 \right>
- \left| 00111001 \right> \\ &&
+ \left| 01010000 \right>
+ \left| 00000101 \right>
+ \left| 01100011 \right>
+ \left| 00110110 \right>
\\
\left| {v011} \right>&=& 
+ \left| 01100000 \right>
+ \left| 00110101 \right>
- \left| 01010011 \right>
- \left| 00000110 \right> \\ &&
- \left| 01101111 \right>
- \left| 00111010 \right>
+ \left| 01011100 \right>
+ \left| 00001001 \right> \\ &&
+ \left| 10011111 \right>
+ \left| 11001010 \right>
- \left| 10101100 \right>
- \left| 11111001 \right> \\ &&
- \left| 10010000 \right>
- \left| 11000101 \right>
+ \left| 10100011 \right>
+ \left| 11110110 \right>
\\ 
\left| {v100} \right>&=& 
+ \left| 10001000 \right>
- \left| 11011101 \right>
- \left| 10111011 \right>
+ \left| 11101110 \right> \\ &&
+ \left| 10000111 \right>
- \left| 11010010 \right>
- \left| 10110100 \right>
+ \left| 11100001 \right> \\ &&
+ \left| 01110111 \right>
- \left| 00100010 \right>
- \left| 01000100 \right>
+ \left| 00010001 \right> \\ &&
+ \left| 01111000 \right>
- \left| 00101101 \right>
- \left| 01001011 \right>
+ \left| 00011110 \right>
\\ 
\left| {v101} \right>&=& 
- \left| 01001000 \right>
+ \left| 00011101 \right>
- \left| 01111011 \right>
+ \left| 00101110 \right> \\ &&
- \left| 01000111 \right>
+ \left| 00010010 \right>
- \left| 01110100 \right>
+ \left| 00100001 \right> \\ &&
- \left| 10110111 \right>
+ \left| 11100010 \right>
- \left| 10000100 \right>
+ \left| 11010001 \right> \\ &&
- \left| 10111000 \right>
+ \left| 11101101 \right>
- \left| 10001011 \right>
+ \left| 11011110 \right>
 \\ 
\left| {v110} \right>&=& 
- \left| 00101000 \right>
+ \left| 01111101 \right>
+ \left| 00011011 \right>
- \left| 01001110 \right> \\ &&
+ \left| 00100111 \right>
- \left| 01110010 \right>
- \left| 00010100 \right>
+ \left| 01000001 \right> \\ &&
- \left| 11010111 \right>
+ \left| 10000010 \right>
+ \left| 11100100 \right>
- \left| 10110001 \right> \\ &&
+ \left| 11011000 \right>
- \left| 10001101 \right>
- \left| 11101011 \right>
+ \left| 10111110 \right>
\\
\left| {v111} \right>&=& 
+ \left| 11101000 \right>
- \left| 10111101 \right>
+ \left| 11011011 \right>
- \left| 10001110 \right> \\ &&
- \left| 11100111 \right>
+ \left| 10110010 \right>
- \left| 11010100 \right>
+ \left| 10000001 \right> \\ &&
+ \left| 00010111 \right>
- \left| 01000010 \right>
+ \left| 00100100 \right>
- \left| 01110001 \right> \\ &&
- \left| 00011000 \right>
+ \left| 01001101 \right>
- \left| 00101011 \right>
+ \left| 01111110 \right>
 \end{eqnarray*}

This code has also recently been derived by Gottesman \cite{Gott96}. He 
presents a general construction for $\{2^r, 2^r-r-2, 3\}$ codes. These 
parameters are consistent with the supposition that such codes are obtained 
from the above method applied to the classical pair $C_2^{\perp} =$ first 
order Reed-Muller $[2^r, r+1, 2^{r-1}]$ code, $C_1 =$ even weight $[2^r, 
2^r-1, 2]$ code. 

Proceeding to the encoding of four qubits, the bound (\ref{bound}) implies 
that single error correction is possible with $n=9$. However, $n=9$ is not 
large enough to allow a significant improvement on the properties of the 
$[8,4,4]$ classical code, so it seems unlikely that $\{9,4,3\}$ is possible, 
and I have not been able to find such a code. With $n=10$, on the other 
hand, we can adopt the 11-bit code indicated in tables \ref{tab1} and 
\ref{tab2} which allows $d=3, d^{\perp}=5$, reducing it by the first 
construction in (\ref{navigate}) to obtain $C_2 = [10,6,3]$, $C_2^{\perp} = 
[10,4,4]$. This leads to the following quantum code: 
  \begin{eqnarray}
G_{\{10,4,3\}} &=& \left( \begin{array}{c}
0101010110 \\
0011001101 \\
0000111100 \\
1111111100 \\
\hline
1100000000 \\
1010000000 \\
1000100000 \\
0000000011   \end{array} \right), \\
S_{\{10,4,3\}} &=& \left( \begin{array}{c}
0011001100110011 \\
0000111100001111 \\
0101010110101010 \\
0000111111110000  \end{array} \right)
\equiv \left( \begin{array}{c}
3333 \\
0{\rm F}0{\rm F} \\
55{\rm AA}  \\
0{\rm FF}0 \end{array} \right).
  \end{eqnarray}
Calderbank {\em et al.} \cite{Cald96b} have also obtained a code of these 
parameters. The quantum Hamming bound (\ref{bound}) does not rule out the 
possibility of a further information qubit without increasing $n$, ie 
$\{10,5,3\}$, but I have been unsuccessful in finding such a code. 

To encode 5 qubits with single error correction, the classical 
11-bit code with $d=3, d^{\perp}=5$ just mentioned can be used to
obtain the following quantum code
  \begin{equation}
G_{\{11,5,3\}} = \left( \begin{array}{c}
01010101100 \\
00110011010 \\
00001111001 \\
11111111000 \\
\hline
11000000000 \\
10100000000 \\
10001000000 \\
00000000110 \\
00000000101  \end{array} \right), \;\;\;\;\;
S_{\{11,5,3\}} = \left( \begin{array}{c}
3333 \\
0{\rm F}0{\rm F} \\
55{\rm AA}  \\
0{\rm FF}0  \\
3333         \end{array} \right).
  \end{equation}
This is optimal for 5 information qubits and single error correction
if $\{10,5,3\}$ does not exist.

\section{Conclusion}

Much research is currently directed to finding the most efficient quantum 
error correction techniques. Commonly in these efforts only the simplest 
example code, encoding a single qubit of information, is actually 
identified. However, to convey many bits of information, it is known from 
classical theory that more advanced codes, involving many information bits, 
are more powerful than a repetition of single-information-bit codes. This 
implies that a more efficient coding technique is not useful unless the task 
of applying it to many information bits is mathematically tractable. A 
simpler coding technique, which is less efficient than other methods for one 
qubit, may become more efficient than the competing methods when many qubits 
are involved, simply because powerful many-qubit codes can be identified for 
the simpler method but not for its competitors. 

In this paper many examples have been given of quantum error correcting 
codes of reasonably high efficiency. In the process, several simple 
techniques for manipulating codes and guessing new ones have been described. 
Quantum networks to encode and correct each code have not been given, since 
they can de deduced directly from the relevent generator and parity check 
matrices\cite{Ste96}[theorem 5]. 

Starting with the simplest general method of quantum error correction, based 
on dual pairs of classical linear codes, and specified here by the notation 
$\{n,K,d\}^+$, we have tabulated codes which can be lifted almost directly 
from classical coding theory because they are self-dual or weakly self-dual. 
In addition, classical dual code pairs with maximal $d, d^{\perp}$ have been 
tabulated, since they form a useful starting point for finding quantum 
codes. We have then examined in more detail the case of single error 
correction, obtaining many good quantum codes from classical codes which are 
not weakly self-dual, and whose conversion to the quantum case therefore 
requires more ingenuity. A method using a subset of the rows of the 
generator matrix as extra parity checks has been described, and used to find 
quantum codes of parameters 
$\{10,2,3\}^+,\;\{12,3,3\}^+$, $\{13,5,3\}^+,\;\{14,6,3\}^+,\; 
\{17,7,3\}^+$, $\{19\cdots 27,8\cdots 16,3\}^+$. 

Next, we have improved on the $\{n,K,d\}^+$ codes by allowing one of the 
classical codes ($C_1$) used to generate the quantum code to have its 
minimum distance reduced, and compensating for this by allowing the signs of 
words in the code vectors to be either positive or negative in all bases. To 
find out how to allocate the signs in this case, it is necessary to use 
insight rather than trial and error. By making the sign allocations 
themselves form a classical linear code, and by using the Hadamard matrix to 
supply useful sets of signs, we have introduced further structure into the 
quantum code. As well as making the design of generator and corrector 
networks easier, this allows the set of possible sign allocations to be 
vastly restricted, which greatly aids the search for good codes. These 
methods have enabled us to identify single-error correcting $\{n,K,3\}$ 
quantum codes of $n=8,10,11$, $K=3,4,5$ qubits respectively. The first is 
optimal, and it is possible that the others are also, though $\{10,5,3\}$ 
and $\{11,6,3\}$ are not ruled out by the quantum Hamming bound. 

Developing general methods for producing good error correcting codes is 
notoriously difficult. The task of finding good codes can be framed as a 
computational problem. It may be an example of a practically important 
computation whose solution on an ideal quantum computer is more efficient 
(has lower computational complexity) than any algorithm for a classical 
computer. 

The author is supported by the Royal Society.


\begin{thebibliography}{12}

\bibitem{Cal96} A. R. Calderbank and P. W. Shor, Good quantum error-correcting
codes exist, submitted to Phys. Rev. A (preprint quant-ph/9512032).

\bibitem{Ste96} A. M. Steane, Multiple particle interference and quantum
error correction, Proc. Roy. Soc. A, to appear (preprint quant-ph/9601029). 

\bibitem{Schu} B. Schumacher, Phys. Rev. A {\bf 51}, 2738 (1995).

\bibitem{Shor} P. W. Shor, Phys. Rev. A {\bf 52}, R2493 (1995).

\bibitem{StePRL} A. M. Steane, Error correcting codes in quantum theory,
submitted to Phys. Rev. Lett.

\bibitem{Laf} R. Laflamme, C. Miquel, J. P. Paz and W. H. Zurek, Perfect 
quantum error correction code, submitted to Phys. Rev. Lett. (preprint 
quant-ph/9602019). 

\bibitem{Ben} C. H. Bennett, D. P. DiVincenzo, J. A. Smolin and
W. K. Wootters, Mixed state entanglement and quantum error correction,
to appear (preprint quant-ph/9604024).

\bibitem{Chuang} I. L. Chuang and R. Laflamme, Quantum Error Correction
by Coding (preprint quant-ph/9511003).

\bibitem{Plenio} M. B. Plenio, V. Vedral and P. L. Knight, Optimal
Realistic Quantum Error Correction Code (preprint quant-ph/9603022).

\bibitem{Ekert96} A. Ekert and C. Macchiavello, Quantum error correction for 
communication, submitted to Phys. Rev. Lett. (preprint quant-ph/9602022). 

\bibitem{MacW} F. J. MacWilliams and N. J. A. Sloane, ``The Theory of
Error-Correcting Codes,'' (North-Holland, Amsterdam 1977). 

\bibitem{HS} H. J. Helgert and R. D. Stinaff, IEEE Trans. Inf. Theory
{\bf 19}, 344 (1973).

\bibitem{ShorSmol} P. Shor and J. A. Smolin, Quantum error-correcting codes
do not need to completely reveal the error syndrome (preprint 
quant-ph/9604006).

\bibitem{Cald96b} A. R. Calderbank, E. M. Rains, N. J. A. Sloane
and P. W. Shor, Quantum error correction and orthogonal geometry
(preprint quant-ph/96).

\bibitem{Gott96} D. Gottesman, A class of quantum error-correcting
codes saturating the quantum Hamming bound (preprint quant-ph/9604038)

\end{thebibliography}
\end{document}